\documentclass[preprint]{IEEEtran}
\IEEEoverridecommandlockouts

\usepackage{cite}
\usepackage{amsmath,amssymb,amsfonts}

\usepackage{graphicx}
\usepackage{epstopdf}
\usepackage{textcomp}
\usepackage{xcolor}

\usepackage{mathtools}
\usepackage{amsmath}
\usepackage{amsthm}
\usepackage{amssymb}
\usepackage{algorithm}
\usepackage{algpseudocode}
\usepackage{accents}

\newcommand\munderbar[1]{%
  \underaccent{\bar}{#1}}

\def\BibTeX{{\rm B\kern-.05em{\sc i\kern-.025em b}\kern-.08em
    T\kern-.1667em\lower.7ex\hbox{E}\kern-.125emX}}
\begin{document}

\title{Collaborative Bearing Estimation Using Set Membership Methods
}
\author{\IEEEauthorblockN{Mohammad Zamani\IEEEauthorrefmark{1},
Jochen Trumpf\IEEEauthorrefmark{2} and
Chris Manzie\IEEEauthorrefmark{3}}
\IEEEauthorblockA{\IEEEauthorrefmark{1}\IEEEauthorrefmark{3}Department of Electrical and Electronic Engineering\\
University of Melbourne,
Melbourne, Australia\\ Email: Mohammad.Zamani@unimelb.edu.au, manziec@unimelb.edu.au}
\IEEEauthorblockA{\IEEEauthorrefmark{2}School of Engineering\\
The Australian National University\\
Canberra, Australia\\
Email: Jochen.Trumpf@anu.edu.au}}

\maketitle

\begin{abstract}
We consider the problem of collaborative bearing estimation using a method with historic roots in set theoretic estimation techniques.  We refer to this method as the Convex Combination Ellipsoid (CCE) method and show that it provides a less conservative covariance estimate than the well known Covariance Intersection (CI) method. The CCE method does not introduce additional uncertainty that was not already present in the prior estimates. Using our proposed approach for collaborative bearing estimation, the nonlinearity of the bearing measurement is captured as an uncertainty ellipsoid thereby avoiding the need for linearization or approximation via sampling procedures. Simulations are undertaken to evaluate the relative performance of the collaborative bearing estimation solution using the proposed (CCE) and typical (CI) methods. 
\end{abstract}

\footnote{Preprint submitted to Fusion 2023}
\section{Introduction}
In this work we revisit the concept of distributed fusion in the context of collaborative networked bearing estimation. Consider a network of estimators each independently estimating the position of a target using information from their onboard bearing sensor and information communicated by their peer nodes in the same network. While the first source of information is assumed to have independent uncertainty, the same cannot be assumed for the second source since information communicated to a given node by a peer node in the present can (and usually will) depend on information that the given node communicated to the peer node in the past, possibly indirectly via a longer communication path in the network. This situation breaks the underlying independence assumption that most fusion algorithms build upon.  Nevertheless, networked estimation is appealing as sharing processed peer information can potentially improve the estimate of each node beyond the information available from their local interactions with the environment. This is particularly true in bearing estimation where \emph{persistency of excitation} from sufficiently diverse bearing directions is necessary for accurate estimation~\cite{Deghat_Bearing}. 

Collaborative networked estimation algorithms have been studied from both top down and bottom up perspectives. Top down approaches start with formulating a joint estimation problem for the entire network followed by decentralisation of the joint estimation algorithm. Often the algorithm can be fully distributed via tightly coordinated communications and assumptions on the network connectivity and reliability. For examples and more details on the top down approach see the discussions in~\cite{roumeliotis2002distributed, roumeliotisCI, luft2018recursive, ZhuRenWeiDistributedLoc}. In contrast, the bottom up approach places less stringent requirements on the network. Bottom up algorithms typically allow for a more opportunistic approach to information sharing that is largely independent of the underlying network topology. However, in order to realise the full potential of this approach we need to address the problem of correlated data when fusing one or more pieces of data over a network, where cross correlations are not tracked as in the top down approaches. 

Algorithms that provide safe fusion in the face of correlated data have been discussed in the literature since at least the 1960s, see for instance~\cite{Schweppe68, Kahan68, BertsekasRhodes71, Fogel79, Fogel1982, Deller89, Sabater02ellcalc, julierUhlmann97CI, Polyak2001}. The proposed solutions are roughly all based on the following principle. If the cross-correlation of information pieces being fused is not known, the fusion process should guarantee to retain the common uncertainty of the original pieces. This way, the fusion cannot result in overconfident estimates across the network. It is generally accepted that optimality of such fusion will be sacrificed for the practicality of the loosely networked approach when compared to the top down methods. The reader is encouraged to consult a more recent paper~\cite{Sabater02ellcalc} for a summary of these classical results and original contributions in terms of explaining the results and offering computationally efficient algorithms to obtain tight and safe fusion.      

Our contribution in this paper is twofold. First, we show that the literature on Covariance Intersection (CI), see e.g.~\cite{julierUhlmann97CI, Chenfusion02, ChenMehra2002estimation,Abu2017,sijs10EI,noack17ICI,wang2022consistentfusion}, including the part of this literature that mentions the connection of CI with the set theoretic (or set membership) results, see e.g.~\cite{ChongMori01, wang2019equivalence}, appears to have mostly overlooked a key aspect of the results in the set theoretic estimation literature. CI and most of the subsequent algorithms in the fusion literature produce an ellipsoidal uncertainty set that is guaranteed to contain the intersection of two uncertainty sets being fused. Furthermore, they offer ways of finding such a set that is minimal either in the determinant or trace over all candidate sets considered by the particular method, a property sometimes referred to as tightness. However, most such methods (including CI) produce more conservative estimates than the set-theoretic method~\cite{Schweppe68, SchweppeBook, Sabater02ellcalc} and, more importantly, do not guarantee that the uncertainty set resulting from the fusion does not introduce any uncertainty not contained in either of the two original sets. We provide a comparison of the uncertainty sets after fusion and also highlight the lack of similar properties for CI~\cite{julierUhlmann97CI} and the more recent Inverse Covariance Intersection (ICI) method~\cite{noack17ICI} with a numerical counter example.

Our second contribution applies the set-theoretic fusion technique to the problem of collaborative bearing estimation. We first introduce a bearing measurement ellipse that models the non-linearity of bearing measurements without requiring linearization or sampling. We show how a Kalman fusion algorithm can be applied to a sequence of such measurements to recursively provide an estimate for the position of the target. The proposed estimation method, while very simple, is very robust to numerical issues and has low computational complexity. We then show how collaborative (and hence potentially correlated) estimates communicated by peer nodes in a network can be incorporated using the set theoretic fusion technique. Lastly we provide a simulation study to highlight the findings of our analysis and to validate our collaborative bearing estimation technique.

\section{Notation}\label{sec:not}
A positive semi-definite (definite) matrix $X\in\mathbb{R}^{n\times n}$ is represented as $X\geq 0 $ ($X>0$). The weighted Euclidean norm of a vector $x\in\mathbb{R}^n$ is denoted by $\Vert x \Vert_X$, $X>0$. 
\begin{equation*}
\Vert x \Vert_X \triangleq \sqrt{x^{\top} X x}. 
\end{equation*}
We use the shorthand notation $\Vert x \Vert:=\Vert x \Vert_I$ for the (unweighted) Euclidean norm.
An ellipsoid $ \mathcal{E}$ in $\mathbb{R}^n$ with center point $c\in \mathbb{R}^n$ and shape matrix $P\in\mathbb{R}^{n\times n}$, $P>0$, is defined as 
\begin{equation*}
\mathcal{E}(c, P) \triangleq \{ x\in \mathbb{R}^n \; : \; \Vert x -c \Vert^2_{P^{-1}}\leq 1\}. 
\end{equation*}

\section{Review of Fusion Algorithms}\label{sec:fusion}

In this section we provide an overview of the so-called Kalman fusion method, the Covariance Intersection (CI) method~\cite{julierUhlmann97CI}, and what we term Convex Combination Ellipsoid (CCE) method from the set-theoretic estimation literature~\cite{Schweppe68, Sabater02ellcalc}. We will show that when we fuse correlated data, as is commonly the case in networked estimation problems, the Kalman fusion method can lead to overconfidence and ultimately large estimation errors (this last point is illustrated in Section~\ref{sec:sims}). CI and the CCE method both avoid the overconfidence problem with CCE shown to yield less conservative covariance estimates than CI. Moreover, we will stress a key property of CCE that seems to have largely been overlooked in the literature, namely that the CCE ellipsoidal estimate is contained within the union of the prior ellipsoidal estimates, potentially explaining the superior performance in our application (cf. Section~\ref{sec:sims}). 

\subsection{Kalman Fusion and the Overconfidence Problem}
The Kalman fusion method can be derived from many different approaches including Bayesian, Fisher and least squares discrete estimation~\cite{SchweppeBook}. Assume that we are given a pair of unbiased estimates $(\hat{x}_i, \hat{P}_i)$ and $(\hat{x}_j, \hat{P}_j)$ for an unknown signal of interest $x\in\mathbb{R}^n$, where $\hat{x}_i, \hat{x}_j\in\mathbb{R}^n$ denote the point estimates and $\hat{P}_i,\hat{P}_j>0$ denote the estimated covariances for their estimation errors. Assuming that these two estimates are \emph{independent} it can be shown, via any of the aforementioned approaches, that the following so called Kalman fusion method provides a fused estimate $(\hat{x}^+,\hat{P}^+)$ that is optimal with respect to many criteria~\cite{sabater1991set}, including the weighted least squares error, minimum covariance estimation error, maximum-likelihood estimation or minimum volume uncertainty ellipsoid. 
\begin{equation}\label{alg:KF}
    \begin{split}
     & \hat{P}^+ = ( \hat{P}_i^{-1} + \hat{P}_j^{-1})^{-1},\\
     & \hat{x}^+ = \hat{P}^+(  \hat{P}_i^{-1}\hat{x}_i + \hat{P}_j^{-1}\hat{x}_j).
    \end{split}
\end{equation}
Here $\hat{x}^+$ denotes the updated point estimate and $\hat{P}^+$ denotes the updated estimation error covariance estimate after fusion. It can be checked that the resulting estimated covariance matrix $\hat{P}^+$ is not larger than the two prior estimated covariances $\hat{P}_i$ and $\hat{P}_j$ due to the parallel summation in the first equation. 

Recall that it was crucially assumed that the two prior estimates $\hat{x}_i$ and $\hat{x}_j$ are independent. If this is not the case,  as in many networked estimation problems where nodes freely pass information back and forth, not only is this formula no longer optimal but it can prematurely result in a small estimated covariance which will hinder the fusion algorithm's ability to correct the point estimate upon receiving any further information. This can be seen in the second equation where the covariance weighted summation of the two prior point estimates is multiplied by the fusion covariance, and hence if one of the prior estimates is overconfident, it will dominate the point estimate of the fusion subsequently. This is referred to as the over-confidence problem in the fusion literature. 

\subsection{Covariance Intersection (CI)}
The Covariance Intersection (CI) method~\cite{julierUhlmann97CI, ChenMehra2002estimation} provides a fused estimate that avoids the over-confidence problem. Specifically, if each prior estimate is conservative in the sense of $\hat{P}_i \geq P_i$, where $P_i$ is the unknown actual covariance of the estimation error $\hat{x}_i - x$, (and similar for $j$) then CI results in an estimate that is also conservative\footnote{This property is often referred to as \emph{consistent} rather than conservative. We avoid the \emph{consistent} terminology since it can be confused with the classical notion of consistency in stochastic estimation~\cite{Jazwinski}.} in the sense of $\hat{P}^+ \geq P^+$, where $P^+$ is now the unknown actual covariance of the estimation error $\hat{x}^+ - x$~\cite{julierUhlmann97CI}. 

The authors in~\cite{julierUhlmann97CI} make the observation that given any known correlation between the two prior estimates, the optimal fusion estimate always has its covariance ellipsoid (i.e. the uncertainty ellipsoid around the origin with covariance as the shape matrix) contained in the intersection of the covariance ellipsoids of the prior estimates.  Motivated by this, they propose the CI method which, by restricting itself to a family of convex combinations of inverse covariance matrices, provides a fusion estimate with its covariance ellipsoid guaranteed to contain the intersection of the two prior covariance ellipsoids (regardless of the actual value of the unknown correlation between the two estimates).  
\begin{equation}\label{alg:CI}
    \begin{split}
     & \hat{P}^+ = (   \alpha \hat{P}_i^{-1} + (1-\alpha)\hat{P}_j^{-1})^{-1},\\
     & \hat{x}^+ = \hat{P}^+(   \alpha \hat{P}_i^{-1}\hat{x}_i + (1-\alpha)\hat{P}_j^{-1}\hat{x}_j).
    \end{split}
\end{equation}
Here, $\alpha \in [0,1]$ is a free parameter. 

\subsection{Convex Combination Ellipsoid (CCE) Fusion}\label{CCE}
 
 An alternative solution for the overconfidence problem is well explained using the notion of uncertainty sets. In this view, an estimator provides an uncertainty set that contains the true signal with certainty or with high probability. Typically, ellipsoids are chosen to represent these sets since they can be simply parametrized in terms of their center and shape matrix, are amenable to matrix calculus, and sufficiently capture the nonlinear nature of complex sets. There are no further assumptions made regarding the distribution of uncertainty inside or outside the uncertainty set in this approach. If we have two or more such uncertainty sets, a fusion uncertainty set is regarded as consistent with these prior estimates if it contains the intersection of the prior uncertainty sets. The exact intersection of prior uncertainty sets (even if they are ellipsoids) is in general difficult to parametrize. Therefore, the tightest (e.g. in the sense of determinant of the shape matrix) ellipsoid that contains the intersection of the prior ellipsoids is taken as the solution. A particular family of sub-optimal solutions to this problem has been known since at least the 1960s~\cite{Schweppe68, Kahan68, SchweppeBook, Sabater02ellcalc}, namely the family of convex combinations of the prior uncertainty sets. This family of solutions not only guarantees tight bounding of the intersection but most importantly also ensures that no uncertainty is introduced that was not already present in the prior uncertainty sets~\cite{Sabater02ellcalc}. While this method is similar in structure to CI, we will show that it has a number of key features which make it an excellent candidate for collaborative bearing estimation.      

 In this approach, node $i$ has an estimate for the unknown signal $x$ parameterised as an ellipsoidal set $\mathcal{E}_i(\hat{x}_i,\hat{P}_i)= \{x\;:\; \Vert x-\hat{x}_i\Vert^2_{\hat{P}_i^{-1}}\leq 1\} $ where $\hat{x}_i$ denotes the center and $\hat{P}_i \geq 0$  denotes the shape matrix (and similar for node $j$).  A stochastic interpretation of this estimate is that the unknown signal $x$ lies in the 1-sub-level set of the function $x\mapsto\Vert x-\hat{x}_i\Vert^2_{\hat{P}_i^{-1}}$ with high probability. Apart from this assertion there is no need to add further assumptions as to how the probability is distributed within or outside of this set. Nevertheless, if we do have a Gaussian interpretation of the errors we can still use this set-theoretic notion as is.

Based on the work presented in~\cite{Sabater02ellcalc, Schweppe68, SchweppeBook}, given two prior ellipsoids whose intersection has nonempty interior, any convex combination of these sets is itself an ellipsoid $\mathcal{E}_{\alpha}(\hat{x}^+,\hat{P}^+)$ given by the following set of equations. We refer to this method as the Convex Combination Ellipsoid (CCE) fusion method. 
\begin{equation}\label{eq:conv_comb}
\begin{split}
&\mathcal{E}_{\alpha}(\hat{x}^+,\hat{P}^+) =\\
&\quad \{x\;:\; \alpha \Vert x-\hat{x}_i\Vert^2_{\hat{P}_i^{-1}} + (1-\alpha)\Vert x-\hat{x}_j\Vert^2_{\hat{P}_j^{-1}}\leq 1\}.\\
\end{split}
\end{equation}
Following some algebra and completing the squares it can be shown that the center and the shape matrix of $\mathcal{E}_{\alpha}$ are as follows.
\begin{equation}\label{alg:CCE}
    \begin{split}
        & \hat{P}^+ = k X,\; 
         X =  (\alpha P_i^{-1} + (1-\alpha)P_j^{-1})^{-1}\\
        & k = 1- d^2,\;
 d^2 = \Vert \hat{x}_j - \hat{x}_i\Vert^2_{(\frac{\hat{P}_i}{\alpha}+\frac{\hat{P}_j}{1-\alpha})^{-1}},\\
        & \hat{x}^+ = X(   \alpha \hat{P}_i^{-1}\hat{x}_i + (1-\alpha)\hat{P}_j^{-1}\hat{x}_j).
    \end{split}
\end{equation}
Note that $k$ is always positive if the two prior ellipsoids have an intersection with nonempty interior and also $k\leq 1$ since clearly $d^2 \geq 0$.

\subsection{Comparison between CI and CCE}\label{sec:comp}
Note that if we fix a value for $\alpha$, both CI~\eqref{alg:CI} and CCE~\eqref{alg:CCE} yield the same point estimate $\hat{x}^+$. The resulting covariance estimate $\hat{P}^+$ of the CI algorithm however is more conservative than that of the CCE algorithm whenever $k<1$, i.e. whenever $\hat{x}_i\not=\hat{x}_j$.

 More importantly, one can show the following properties for CCE~\cite{Sabater02ellcalc}:
\begin{itemize}
    \item The ellipsoid $ \mathcal{E}_{\alpha}$ contains the intersection of the two prior ellipsoids, $\mathcal{E}_i\cap \mathcal{E}_j\subseteq \mathcal{E}_{\alpha}$.
    \item The intersection of the boundaries of $\mathcal{E}_i$ and $\mathcal{E}_j$ is on the boundary of $ \mathcal{E}_{\alpha}$.
    \item The ellipsoid $ \mathcal{E}_{\alpha}$ is contained in the union of the two prior ellipsoids, $ \mathcal{E}_{\alpha}\subseteq  \mathcal{E}_i \cup \mathcal{E}_j$.
    
\end{itemize}

As the authors in~\cite{Sabater02ellcalc} point out, the first two properties ensure tight bounding of the intersection and the third property ensures that no additional uncertainty is introduced that was not already present in the prior uncertainty sets. This last property can play a key role in many applications and seems to have been largely overlooked, even in papers where the CCE method has been mentioned in comparison to CI, e.g. in~\cite{ChongMori01, wang2019equivalence}. 

In both the CI and CCE algorithms one can choose the optimal solution in their respective families of solutions that are parametrized by $\alpha$, for example such that the resulting covariance estimate is minimal in trace or determinant. The latter criterion is proportional to the volume of the covariance ellipsoid while the former equals the sum of squared eigenvalues of the covariance estimate. These optimisation problems have been shown to be convex scalar problems~\cite[Lemma 2]{Sabater02ellcalc} that can either be solved with off-the-shelf solvers, with simple line search or various other heuristics. Optimising $\alpha$ for the CCE algorithm will incur slightly more computation than for CI, but even in the case of CCE it was shown in~\cite{Sabater02ellcalc} that one can reduce the problem to solving a polynomial equation. More precisely, $\alpha^*$ for CCE can be found as the only root in $(0,1)$ of a polynomial of degree $2n-1$. The algorithm is efficient, requires no matrix inversion and is numerically robust.   

\begin{figure}
    \centering
    \includegraphics[width=.5\textwidth]{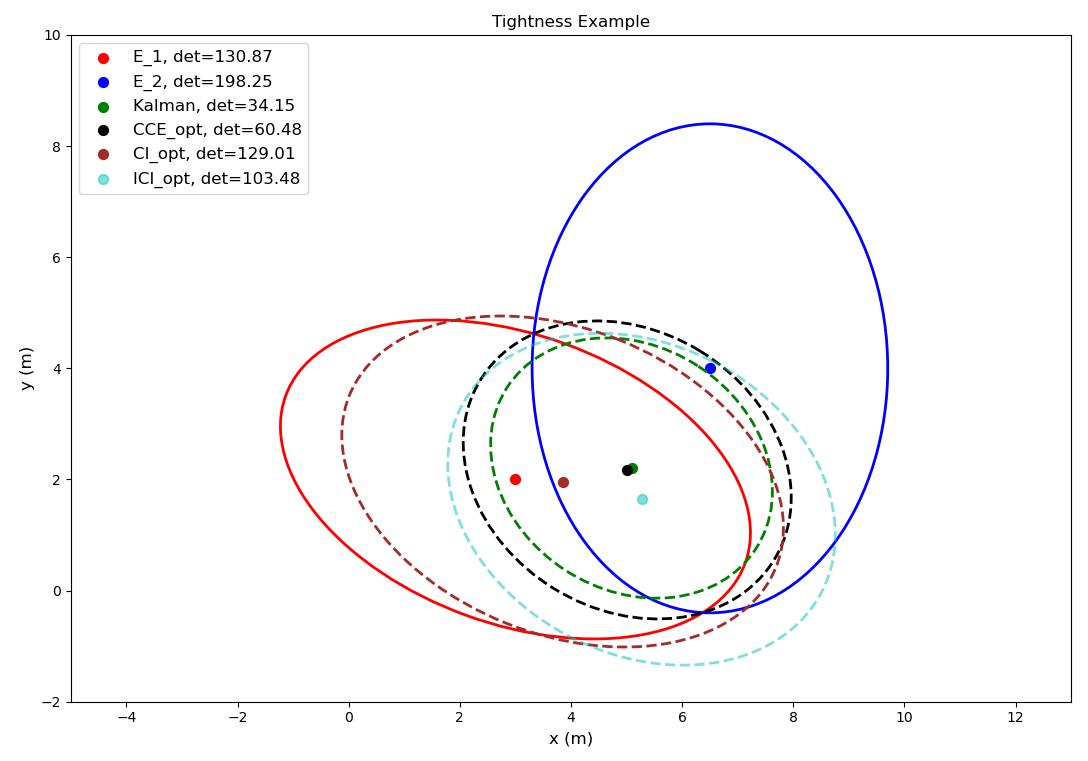}
    \caption{A counterexample where (optimal) CI~\cite{julierUhlmann97CI, ChenMehra2002estimation} and (optimal) ICI~\cite{noack17ICI} violate some of the CCE properties discussed in Section~\ref{sec:comp}. This in particular also shows that the covariance ellipsoid produced by ICI, while having a smaller determinant compared to CI, does not always contain the full intersection of the prior ellipsoids.}
    \label{fig:comp}
\end{figure}

We provide an example in Figure~\ref{fig:comp} where it can be seen that the second and third property are not generally guaranteed for CI (the first property holds for CI), not even in the optimal determinant case. For the Inverse Covariance Intersection method~\cite{noack17ICI}, a more recent development in the fusion literature, none of the three properties can be guaranteed as the same example shows, again not even in the optimal determinant case. 

\section{Problem Formulation and Solution Approach}\label{sec:problem}
Consider the problem of estimating a target position $p\in\mathbb{R}^n$ collaboratively using $m$ agents located at $x_i\in\mathbb{R}^n$, $i=1,\dots,m$. In order to simplify the exposition we will consider $n=m=2$ but the results we present can easily be extended to higher dimensional state spaces and more agents.

Let us assume agent $i$, using local measurements and information from its neighbour $j$, calculates $\hat{x}_i\in\mathbb{R}^n$ as a point estimate for the target position $p_i$, and $\hat{P}_i\in\mathbb{R}^{n\times n}$ as the estimate of its estimation error covariance $P_i\in\mathbb{R}^{n\times n}$.

An outline of our proposed approach is provided in Algorithm~\ref{alg:proposed}, and the following subsections detail the key steps.

\subsection{Measurement Model}\label{sec:meas}
Each agent is equipped with a bearing sensor producing bearing angle measurements $\theta_i$ in $\mathcal{S}^1$.
\begin{equation}\label{eq:meas}
    \theta_i = \arctan{\dfrac{p -x_i}{\Vert p -x_i\Vert}} + \delta_i,   
\end{equation}
where $\delta_i\in\mathcal{S}^1$ is an unknown measurement error with the assumption that $\delta_i\sim \mathcal{N}(0,\sigma_i^2)$ with $\sigma_i$ a known standard deviation. 

Equation~\eqref{eq:meas} is nonlinear in $p$ and will either need linearization for the Kalman filter to be applicable or can be treated using a nonlinear filtering approach. Here, we employ an ellipsoidal measurement modelling approach instead.

\begin{figure}[h]
    \centering
    \includegraphics[width=.5\textwidth]{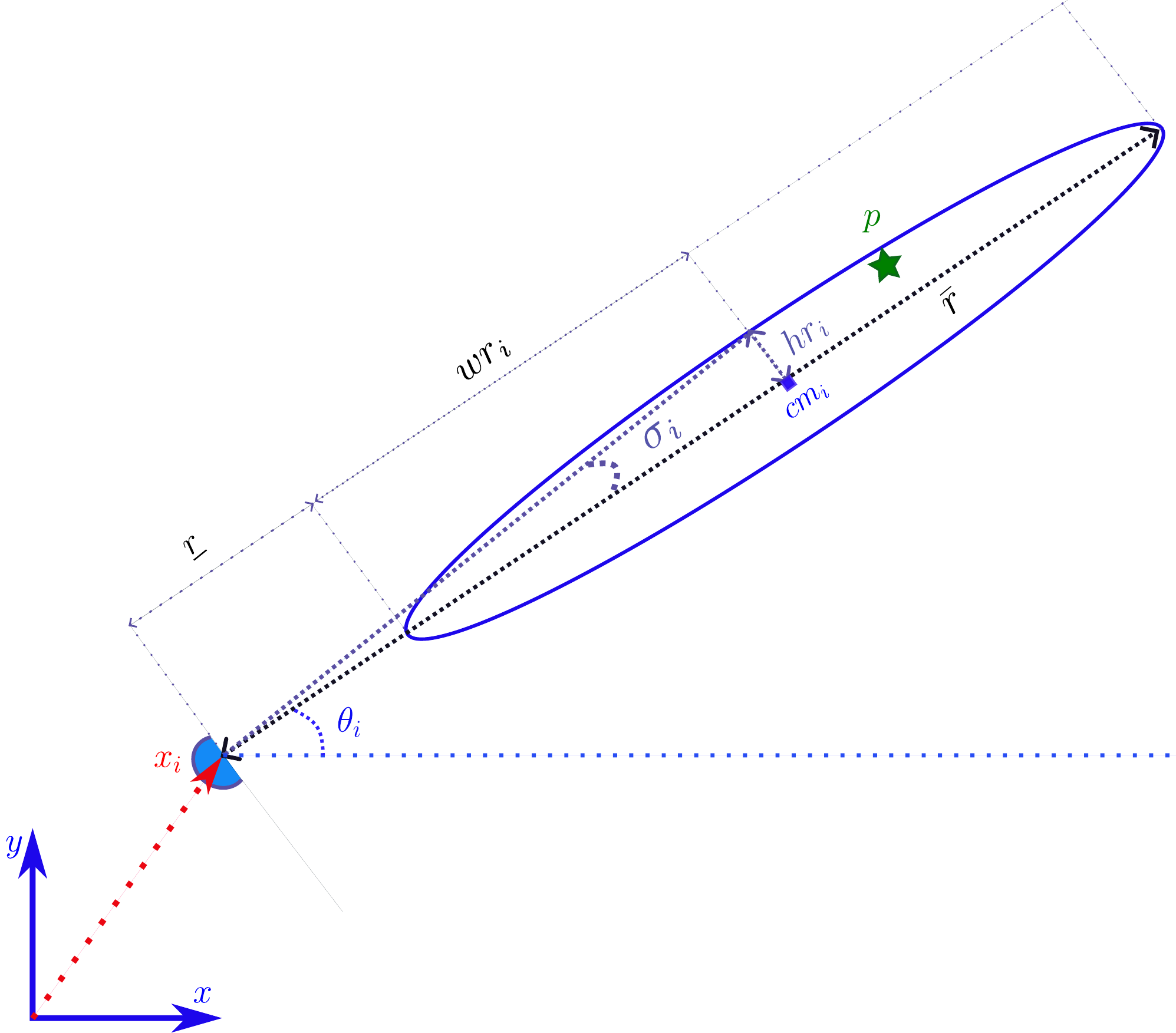}
    \caption{The measurement ellipse for the sensor of agent $i$ located at $x_i$ measuring a noisy bearing angle $\theta_i$ towards a target located at $p$. It is  assumed that the sensor has a measurement range in $[\underbar{$r$}_i, \bar{r}_i]$ and a bearing standard deviation error of $\sigma _i$.}
    \label{fig:meas_ellipse}
\end{figure}

Assume further that we know a minimum and maximum range, $\munderbar{r}_i$ and $\bar{r}_i$ respectively, within which our bearing sensor is designed to operate. Then, we can calculate a  measurement ellipse $\mathcal{E}^m_i(cm_i, Pm_i)$ with center $cm_i$ and covariance matrix $Pm_i$ for the 2D case, $n=2$. A similar approach can be extended to 3D as well but for simplicity of exposition we keep to 2D. 
\begin{equation}\label{eq:meas_ellipse1}
\begin{split}
    &cm_i = x_i + \big(\dfrac{\munderbar{r}_i+\bar{r}_i}{2}\big)\begin{bmatrix}
        \cos{\theta_i}\\\sin{\theta_i}
    \end{bmatrix}\\
    &Pm_i = R D R^{\top},
\end{split}
\end{equation}
where
\begin{equation}\label{eq:meas_ellipse2}
\begin{split}
& wr_i = \dfrac{\bar{r}_i-\munderbar{r}_i}{2}, \; hr_i = \big(\dfrac{\munderbar{r}_i+\bar{r}_i}{2}\big)\tan{\sigma_i},\\
&
    R_i = \begin{bmatrix}\cos{\theta_i} & -\sin{\theta_i}\\
    \sin{\theta_i} & \cos{\theta_i}\end{bmatrix},\; D= \begin{bmatrix}wr_i^{2} & 0 \\
    0 & hr_i^{2}\end{bmatrix}. 
    \end{split}
\end{equation}
These equations are geometrically represented in Figure~\ref{fig:meas_ellipse}.
This model captures the fact that a bearing sensor provides little depth information and hence the measurement ellipse will be stretched longer in the direction towards the target compared to the perpendicular direction. 
\subsection{Estimation and Fusion}
\begin{algorithm}[hbt!]
\caption{Collaborative Bearing Estimation}\label{alg:proposed}
\begin{algorithmic}
\Require initial guess $(\hat{x}_i, \hat{P}_i)$, $\hat{P}_i>0$. 
\Require sensor parameters $\munderbar{r}_i, \bar{r}_i$ and $\sigma_i$.

\While{estimating target}
\State Broadcast $(\hat{x}_i, \hat{P}_i)$ to other nodes on the network.
\If{new measurement $\theta_i$ is received}
    \State Calculate an uncertainty ellipse $\mathcal{E}^m_i(cm_i, Pm_i)$ (Section~\ref{sec:meas}). 
    \State Check if this set overlaps $\mathcal{E}_i(\hat{x}_i, \hat{P}_i)$ (Section~\ref{sec:overlap}).
    \If{there is overlap}
    \State Calculate $\hat{x}^+_i$ and $\hat{P}_i^+$ using the Kalman fusion method~\eqref{alg:KF}.
    \Else{}
    \State Discount the measurement, $Pm_i \gets d_m Pm_i$ (Section~\ref{sec:overlap}).
    \State  Calculate $\hat{x}^+_i$ and $\hat{P}_i^+$ using the Kalman fusion method~\eqref{alg:KF}.
    \EndIf
    \State $(\hat{x}_i, \hat{P}_i) \gets  (\hat{x}^+_i, \hat{P}^+_i)$.  
\ElsIf{communication $\mathcal{E}_j(\hat{x}_j, \hat{P}_j)$ is received}
    \State Check if this set overlaps $\mathcal{E}_i(\hat{x}_i, \hat{P}_i)$ (Section~\ref{sec:overlap}).
    \If{there is overlap}
    \State Calculate $\hat{x}^+_i$ and $\hat{P}_i^+$ using the CCE method~\eqref{alg:CCE}.
    \Else{}
    \State Discard the communication. 
    \EndIf
    \State $(\hat{x}_i, \hat{P}_i) \gets  (\hat{x}^+_i, \hat{P}^+_i)$.
\EndIf
\EndWhile
\end{algorithmic}
\end{algorithm}
We seek to solve the following problem. Each node $i$ calculates an estimate $\hat{x}_i(t)$ for the unknown position of the target $p(t)$, and $\hat{P}_i(t)$ for the covariance matrix characterising the uncertainty of the estimation error.

A recursive solution is sought such that at each time step $t$ the estimate $(\hat{x}_i(t), \hat{P}_i(t))$ is calculated based only on the previously calculated estimate $(\hat{x}_i(t-1), \hat{P}_i(t-1))$ and the new measurements or communications received at time $t$. 

At time $t=0$, $(\hat{x}_i(0), \hat{P}_i(0))$ can be based on a prior (informed) guess. We propose to interpret $(\hat{x}_i(t), \hat{P}_i(t))$ as the center and the shape matrix that characterises an uncertainty set for the target position in the sense that the target lies within this uncertainty set with a high probability. Recall also the local measurement model proposed in~\eqref{eq:meas} which provides another such uncertainty set with its center and shape matrix given in~\eqref{eq:meas_ellipse1} and~\eqref{eq:meas_ellipse2}, respectively. When agent $i$ obtains a measurement (i.e. when the target is within the range $[\munderbar{r}_i,\bar{r}_i]$ of the bearing sensor of agent $i$) we propose to use the Kalman fusion algorithm~\eqref{alg:KF} to update the estimate of agent $i$. This is reasonable since each new sensor measurement can be expected to be independent of the current estimate.  When another agent $j$ shares its current estimate with agent $i$ it too specifies an uncertainty set for the target. In this case we can use one of the Kalman fusion, CI, ICI or CCE methods in order to update the current estimate of the agent. As was discussed before, in this case one cannot assume independence of the information that is communicated across a network, and we propose that the CCE method is the best choice for this problem.

\subsection{Dealing with Non-Overlapping Sets}\label{sec:overlap}
A potential challenge that can occur in using any of the fusion methods of Section~\ref{sec:fusion}  is that the prior uncertainty ellipsoids might not overlap. A heuristic method based on the Mahalonobis distance is proposed here to deal with this issue.

First we verify whether the to be fused ellipsoid has an overlap with the current estimated uncertainty ellipsoid. This can be easily checked using the Mahalanobis distance, shown here for the case of fusing a measurement ellipsoid.
\begin{equation}
    d_m = \Vert cm_i - \hat{x}_i \Vert _{(\hat{P}_i + Pm_i)^{-1}}. 
\end{equation}
If $d_m \leq 2$ we can assert that the two ellipsoids are indeed overlapping. In this case the Kalman fusion method~\eqref{alg:KF} is used to merge the two ellipsoids into one. If $d_m > 2$ the two ellipsoids are disjoint. In this case we can either proceed with the same Kalman fusion approach or we can discard (or discount) one of the two ellipsoids (inversely proportional to $d_m$). We propose to discount the measurement ellipsoid in the case of disjointedness. Again this is on the basis that each new measurement is independent of the current estimate and hence it may include corrective information that should not be fully discarded.   

When information is obtained from a neighboring agent, we can again use the Mahalanobis distance to check disjointedness of the two estimates. We recommend discarding the communicated estimate if disjointedness is detected as there is no reason why the communicated information should be preferred to the current estimate, and it is unclear how to compute reasonable weights for both pieces of information in this case. 

\section{Simulation Study}\label{sec:sims}
In this section we provide a numerical evaluation of the method proposed in Section~\ref{sec:problem}.

\begin{figure}
    \centering
    \includegraphics[width=.5\textwidth]{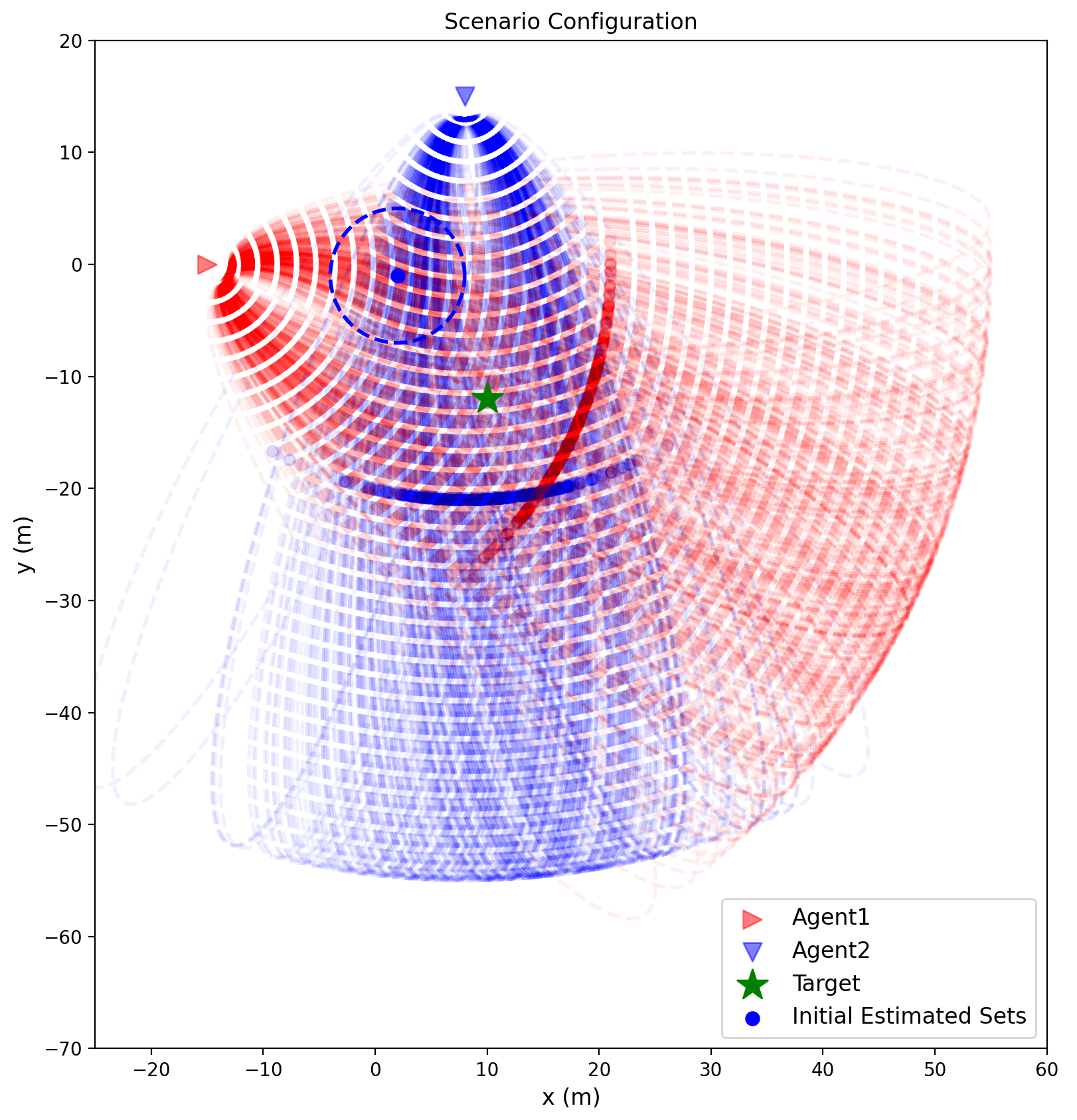}
    \caption{The 2-agent bearing estimation scenario where the initial estimated uncertainty sets are identical circles centered at $[2,-1]^{\top}$. The noisy measurement ellipses of each agent are depicted in the color corresponding to the agent.}
    \label{fig:scenario}
\end{figure}

Consider a stationary target estimation scenario where two stationary agents collaboratively estimate the 2D position of the target. The positioning of the agents is such that each agent individually is unable to obtain a good target position estimate. This setup is depicted in Figure~\ref{fig:scenario}. As can be seen from the figure, Agent 1 is geometrically best positioned to estimate the y-coordinate of the target while Agent 2 is best positioned for estimating the x-coordinate of the target. The noisy measurement ellipses of each agent are also depicted for all the measurements made by the agents during the entire simulation. The agents start from the same initial estimated uncertainty set as depicted in the figure. The detailed parameters of this study are as follows.

The target, Agent 1 and Agent 2 are located at $p = [10, -12]^{\top}$, $x_1 = [-15, -0]^{\top}$ and $x_2=[8, 15]^{\top}$, respectively. A Monte Carlo simulation is conducted with randomised initial estimates, initial estimated covariances and standard deviations of bearing measurements.
Each simulation instance runs for 300 steps.

\begin{figure}
    \centering
    \includegraphics[width=.5\textwidth]{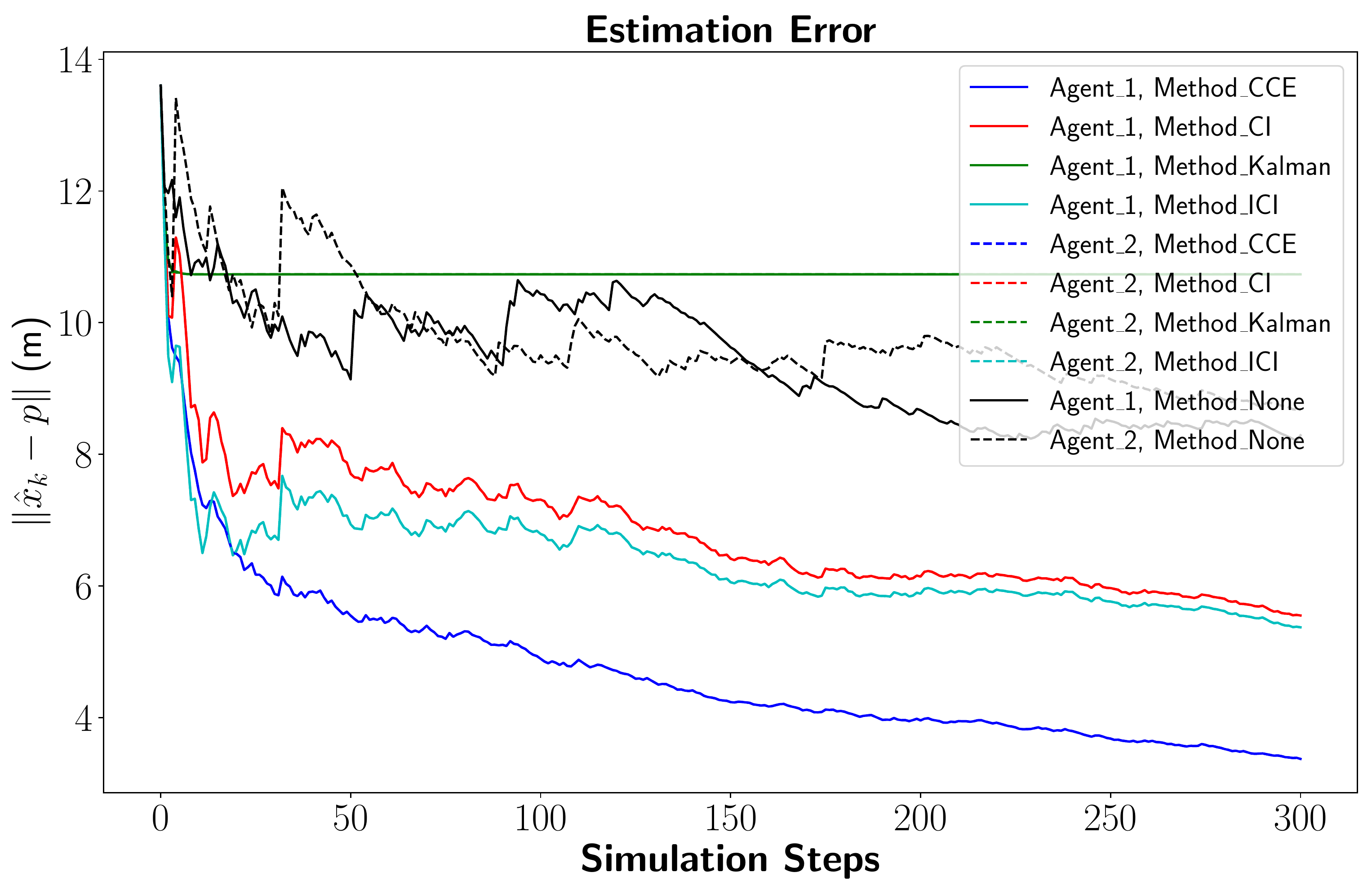}
    \caption{The time trajectory of the estimation error for the non-collaborative, Kalman, CI, ICI, and CCE methods for each agent.}
    \label{fig:est_err}
\end{figure}

\begin{figure}
    \centering
    \includegraphics[width=.5\textwidth]{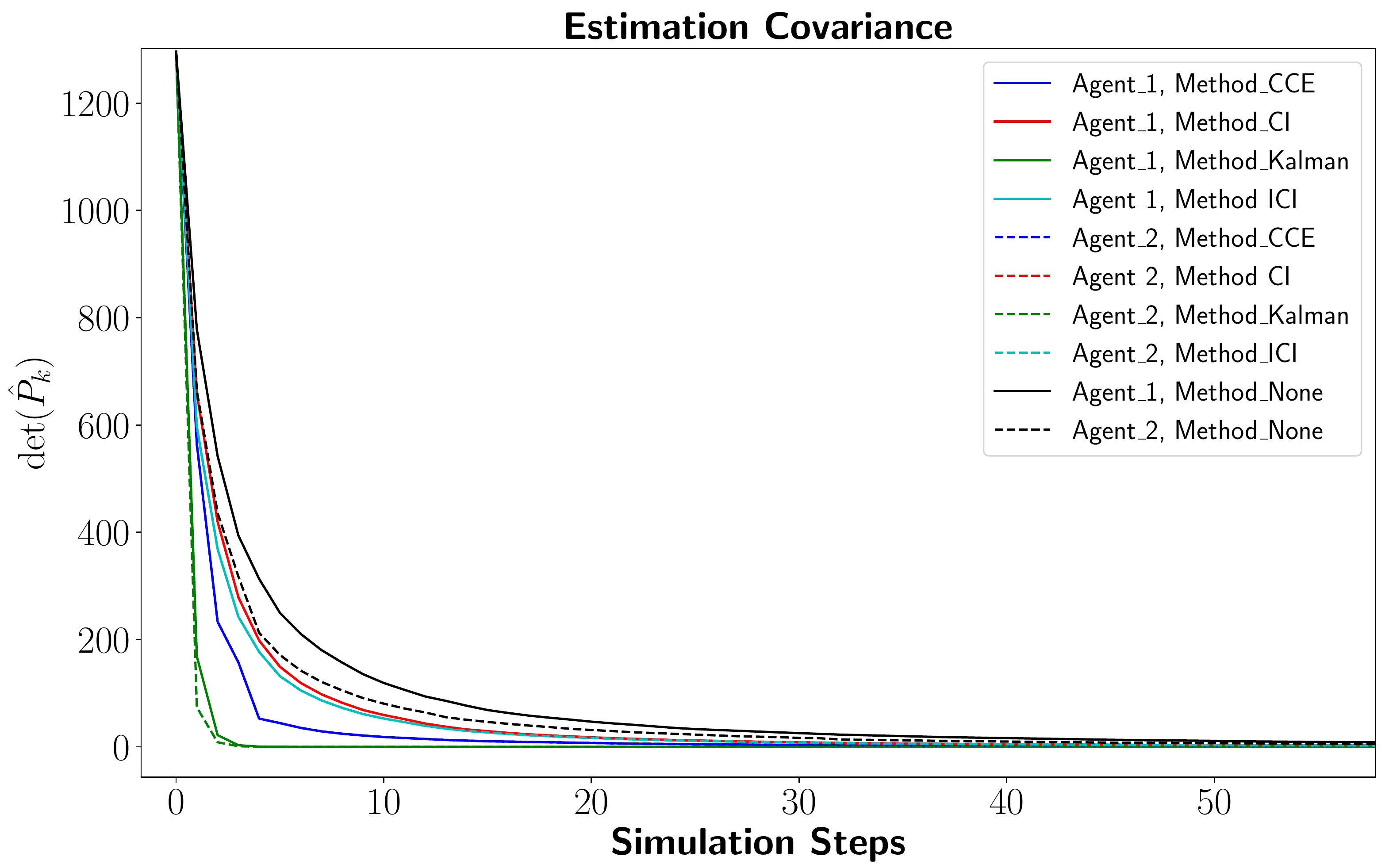}
    \caption{The time trajectory of the determinant of the estimated error covariance is shown for the non-collaborative, Kalman, CI, ICI and CCE methods for each agent. Note that the figure is zoomed to the first 50 simulation steps.}
    \label{fig:covariance}
\end{figure}

We first discuss one instance of the simulation, showing a typical outcome. This allows us to inspect the time trajectories for this particular instance.
The initial estimates for this instance are $\hat{x}_1=\hat{x}_2=[2, -1]^{\top}$, $\hat{P}_1=\hat{P}_2=36 I_{2\times 2}$. The measurement ellipse parameters are $\munderbar{r}_1=\munderbar{r}_2= 2$ (m), $\bar{r}_1=\bar{r}_2=70$ (m), $\sigma_1 = 12$ (deg) and $\sigma_2 = 10$ (deg). The estimation error, $\Vert \hat{x}_i - p\Vert$, and the determinant of the estimated covariance,  $\det{\hat{P}_i}$, for each method and for both agents $i=1,2$ is shown in Figures~\ref{fig:est_err} and~\ref{fig:covariance}, respectively.       

Figure~\ref{fig:est_err} shows that that the CCE method outperforms ICI, CI, Kalman fusion and the non-collaborative methods. It can be seen that Kalman fusion quickly becomes over-confident, stops accepting new measurement information and converges to a final estimation error of $10$m starting from an initial estimation error of $14$m. The non-collaborative agents can only reduce their errors close to $6$m and $9$m, respectively. This is expected since the scenario is designed such that each agent can mainly improve on only one coordinate based on its own measurements. The ICI method slightly outperforms the CI method but both result in just under $6$m of estimation error while the CCE method achieves a final estimation error of $4$m, out-performing the other methods. It can be seen in Figure~\ref{fig:covariance} that the estimated covariance of the Kalman fusion method converges very quickly due to the underlying (erroneous) independence assumption for the information obtained from communications. The other methods are more conservative but CCE produces tighter (more confident) results than both ICI and CI. 

\begin{figure}
    \centering
    \includegraphics[width=.5\textwidth]{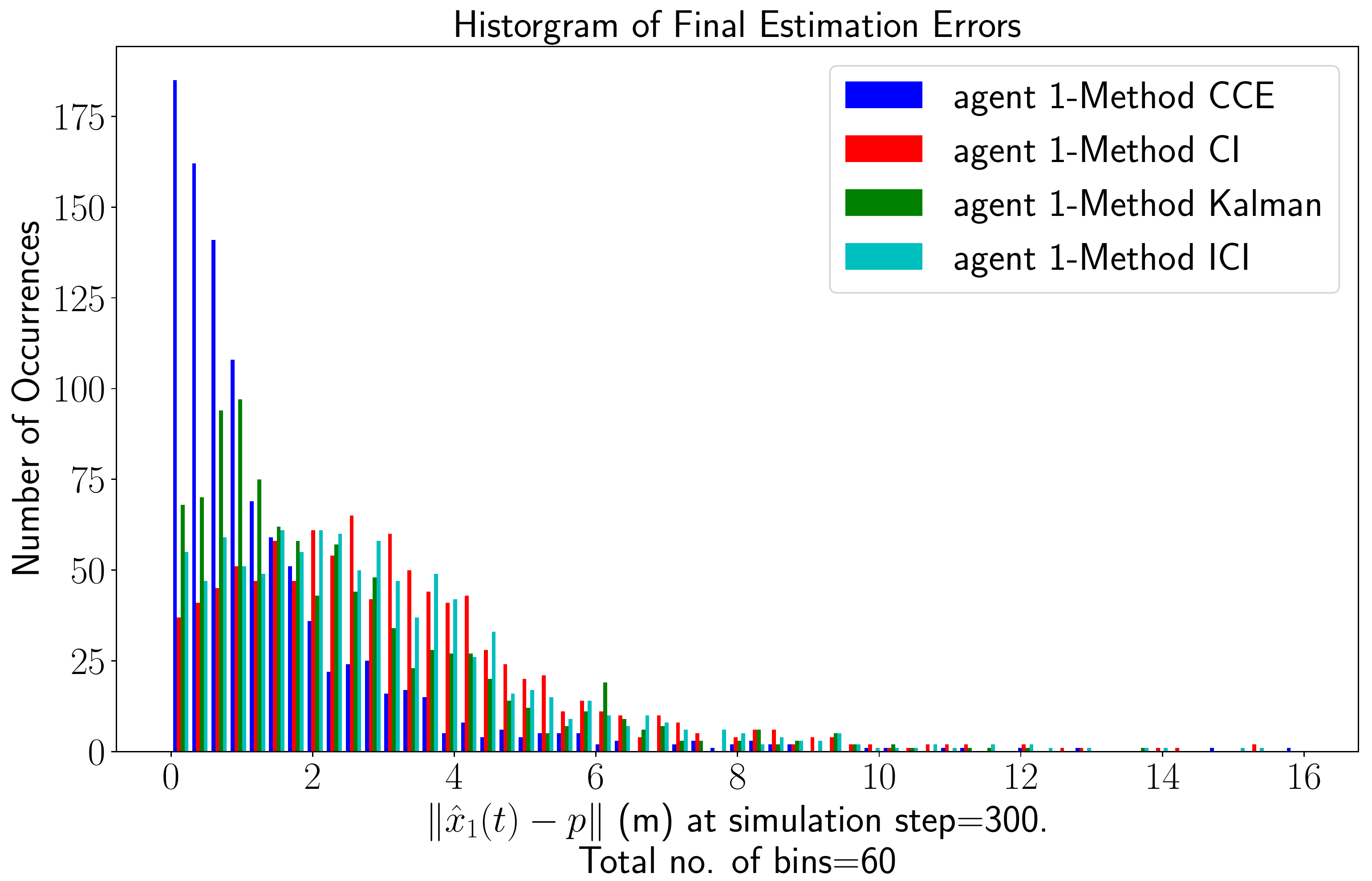}
    \caption{Histogram the final estimation error of the Kalman, CI, ICI and CCE methods at the final simulation step. The Monte Carlo experiment involves 1000 simulation runs where initial estimates, sensor ranges and bearing errors are randomised.}
    \label{fig:hist}
\end{figure}

In the Monte Carlo setup we use the following randomisation and run $1000$ experiments.
Consider two random standard deviations as $\gamma_1,\gamma_2\sim \mathcal{N}(10,100)$, and two error vectors $e_1 \sim\mathcal{N}([0,0]^{\top},[\gamma_1^2,\gamma_1^2]^{\top})$ and $e_2 \sim\mathcal{N}([0,0]^{\top},[\gamma_2^2,\gamma_2^2]^{\top})$. The initial estimates are $\hat{x}_1=p + e_1$ and $\hat{x}_2=p + e_2$. The associated initial covariances are $\hat{P}_1=\gamma_1^2 I_{2\times 2}$ and $\hat{P}_2=\gamma_2^2 I_{2\times 2}$. The measurement ellipse parameters are randomly drawn as  $\munderbar{r}_1,\munderbar{r}_2 \sim \mathcal{N}(2,25)$ (m), $\bar{r}_1,\bar{r}_2\sim \mathcal{N}(80,400)$ (m) and $\sigma_1,\sigma_2 \sim \mathcal{N}(5,25)$ (deg). Figure~\ref{fig:hist} summarises the results of the Monte Carlo simulation in a histogram plot of the final estimation error of Agent 1. The final errors for Agent 2 are similar and have been omitted to simplify this figure. Again, it can be seen that CCE significantly outperforms the other methods in the majority of simulation runs.

\section{Conclusions}
\label{sec:conclusions}
In this paper we propose a collaborative target estimation problem where agents complement their local measurements with fusion of estimates communicated by neighboring agents. We propose a nonlinear model for bearing measurements that can directly feed into the discrete Kalman filter without requiring linearization or sampling-based approximations. The filter remains numerically robust using this measurement model. We propose to pair this measurement model with a lesser known method of fusion from the set membership literature that we call Convex Combination Ellipsoid (CCE) fusion. We discuss how this method addresses the overconfidence problem resulting from correlated network information. Furthermore, we stress key properties of this method that are not shared by the well known CI method, including improved tightness and avoiding new uncertainty as byproduct of the fusion process.     

\bibliographystyle{IEEEtran}
\bibliography{ref}

\end{document}